\documentclass[twocolumn,aps,10pt,showpacs,showkeys,]{revtex4}
\usepackage[dvips]{graphicx}

\begin{document}
\title{Comment on a Phys. Rev. Lett. paper: 94 (2005) 146402:\\
Orbital symmetry and Electron Correlation in Na$_{x}$CoO$_{2}$}
\author{R. J. Radwanski}
\homepage{http://www.css-physics.edu.pl}
\email{sfradwan@cyf-kr.edu.pl}
\affiliation{Center of Solid State Physics, S$^{nt}$Filip 5, 31-150 Krakow, Poland,\\
Institute of Physics, Pedagogical University, 30-084 Krakow,
Poland}
\author{Z. Ropka}
\affiliation{Center of Solid State Physics, S$^{nt}$Filip 5,
31-150 Krakow, Poland}

\begin{abstract}
We argue that the electronic structure considered in a Phys. Rev.
Lett. paper 94 (2005) 146402 of the Co$^{3+}$ ion in the CoO$_{6}$
octahedron of Na$_{x}$CoO$_{2}$ is completely wrong. The
presented Fig. 1 is redrawn here as Fig. 1. For physically
adequate electronic structure it is necessary to take into
account strong intra-atomic electron correlations and spin-orbit
coupling. For Co$^{3+}$ ions there are 15 low lying many-electron
states within 0.1 eV as can be obtained in the many-electron CEF
approach.

\end{abstract}
\maketitle

We claim that the electronic structure considered in a Phys. Rev.
Lett. paper of Wu et al. \cite{1} for the Co ion in the CoO$_{6}$
octahedron in Na$_{x}$CoO$_{2}$, Fig. 1b, is completely wrong.
The presented Fig. 1 in the commented paper is redrawn here as
Fig. 1 together with the caption.

An attributing of schematic 5 lines in Fig. 1b to the ionic model
we treat as a fun. It is a nonsense, but not an ionic model.

We write this Comment as this nonsense is almost typical for Phys.
Rev. journals (not mentioning others). Even if the shown states
could have some sense for one d electron these electronic states
are totally wrong for the Co ion, with 5 (Co$^{4+}$), 6
(Co$^{3+}$) or 7 (Co$^{2+}$) $d$ electrons, in the CoO$_{6}$
octahedron as is described in the figure caption.

Despite of such strong critics of the theoretical introduction we
like the commented paper, in particular its clear conclusion,
that their XAS experiment provides "spectral evidence for strong
electron correlations of the layered cobaltates". The existence
of strong electron correlations in real 3$d$-ion compounds is
exactly this a physical fact that causes the erroneous of Fig.
1b. The necessity of taking into account strong electron
correlations in description of 3$d$-ion compounds we point out
from at least 10 years discussing 3$d$-ion compounds within the
(many-electron) crystal-field (CEF) approach extended to a Quantum
Atomistic Solid State Theory (QUASST) \cite{2,3}. Just the
presence in the reality of these strong intra-atomic correlations
make the single-electron picture, presented in the commented
paper, completely inadequate for the cobalt ion.

Before further discussion we would like to say that advocating
for the fundamental importance of the CEF approach in combination
with the spin-orbit coupling we are aware that all exotic
phenomena existing in 3$d$-ion compounds, in particular in
Na$_{x}$CoO$_{2}$ cannot be explained in the purely CEF approach.
For explaining our point let concentrate on NaCoO$_{2}$, i.e. x=1
case. In this case for the good stoichiometry sample we expect to
have all Co ions in the trivalent state, noted as the Co$^{3+}$
ion. For x$<$1 sample some Co ions are forced to be in the
tetravalent state and consequently we have a complex mixed system
that can show new phenomena (Fermi level, itineracy related to a
geometrical charge order/disorder, ...). Thus, let concentrate on
perfect NaCoO$_{2}$ with only Co$^{3+}$ ions.

\begin{figure}[t]
\begin{center}
\includegraphics[width = 8 cm]{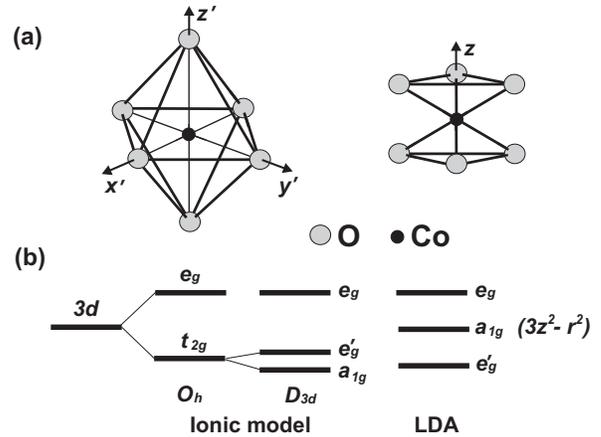}
\end{center} \vspace {-0.3 cm}
\caption{(a) Illustration of the trigonal distortion of a
CoO$_{6}$ octahedron. Left panel: undistorted CoO$_{6}$ octahedron
with cubic ($O_{h}$) symmetry. Right panel: compressed CoO$_{6}$
octahedron with $D_{3d}$ symmetry. The distorted CoO$_{6}$ is
rotated such that the threefold rotation axis is along the $c$
axis. (b) Crystal-field splitting of Co 3$d$ states in distorted
CoO$_{6}$ according to an ionic model and relative energy
positions of 3$d$ bands obtained from LDA calculations. Redrawn
from Ref. \cite{1} together with the full caption.}
\end{figure}

\begin{figure}[ht]
\includegraphics[width = 7.0 cm]{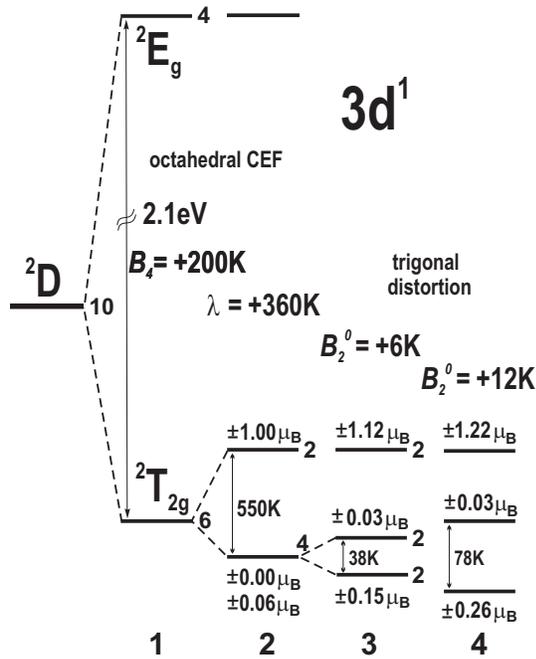}
\caption{Calculated localized states of the 3$d^{1}$ configuration in the V$%
^{4+}$ ion under the action of the crystal field and spin-orbit
interactions originated from the 10-fold degenerated $^{2}D$
term; (1) the splitting of
the $^{2}D$ term by the octahedral CEF surroundings with $B_{4}$=+200 K, $%
\lambda _{s-o}$ =0; (2) the splitting by the combined octahedral
CEF and spin-orbit interactions; (3) and (4) the effect of the
trigonal distortions. The states are labeled by the degeneracy in
the spin-orbital space and the value of the magnetic moment.
Redrawn from our paper, Ref. \cite{9}. For the Ti$^{3+}$ ion the
spin-orbit splitting is smaller due to smaller value of the
spin-orbit coupling. In our paper \cite{5} we used
$\lambda_{s-o}$ = +220~K that yields the s-o splitting of 330 K. }
\end{figure}

\begin{figure}[t]
\begin{center}
\includegraphics[width = 7.3 cm]{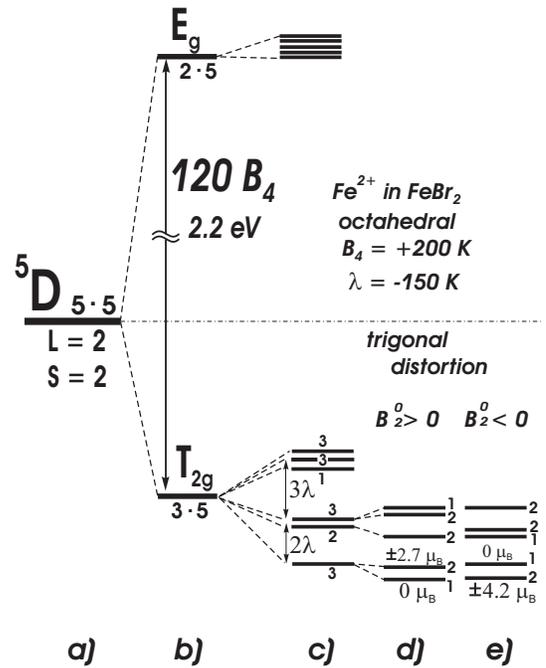}
\end{center}\vspace {-0.7 cm}
\caption{The fine electronic structure of the highly-correlated
3$d^6$ electronic system. a) the 25-fold degenerated $^5D$ term
given by Hund's rules: S=2 and L=2. b) the effect of the cubic
octahedral crystal-field, c)
the combined action of the spin-orbit coupling and the cubic crystal field: B%
$_4$=+200K, $\lambda $= -150 K; d and e) an extra splitting
produced by the trigonal distortion - the case (e), trigonally
stretched octahedron, is realized in FeBr$_2$; the
trigonal-distortion parameter B$_2^0$= -30~K produces a spin-like
gap of 0.0028 meV \cite{13}. The case (d) occurs for the
trigonally compressed octahedron, i.e. exactly like in Fig. 1b
($D_{3d}$) of the commented paper. }
\end{figure}
In an ionic model that we propagate by last years (we do not
claim to invent the crystal-field theory, as it was invented in
1929-1932 by Bethe, Kramers, Van Vleck and many others, but in
last 20 years we propagate the CEF approach being continuously
discriminated scientifically; this discrimination with some
inquisition incidents of Phys. Rev. Editors (M. Blume, Wells, ..)
and of Polish scientific institutions with a help of Prof. H.
Szymczak, J. Sznajd, J. Klamut, A. M. Oles, entitles us to feel at
present ourselves as reinventors of the crystal-field in the
solid-state physics \cite{4}, the more that we extend it from the
single-ion theory to the solid-state theory pointing out for
3d-ion compounds, for instance, the importance of the spin-orbit
coupling \cite{5} and of local distortions) we assume on-site
electron correlations to be sufficiently strong to keep the
atomic-like integrity of the 3$d$ ion. It means, that we consider
that a 3$d$ atom becoming a full part of a solid preserves
largely its individuality and its 3$d$ electrons form a
strongly-correlated electron system 3$d^{n}$, here 3$d^{6}$
system for the Co$^{3+}$ ion. The realization of the specific
valency, i.e. the specific electron configuration in a compound,
depends on the stoichiometry and partners, i.e. Co$^{3+}$ with
the 3$d^{6}$ configuration occurs in LaCoO$_{3}$ \cite{6},
whereas in CoO occurs the Co$^{2+}$ ion with the 3$d^{7}$
configuration \cite{7}.

Before describing the really-ionic states of the 3$d^{6}$ system
we present electronic states for the 3$d^{1}$ configuration, Fig.
2. Such the 3$d^{1}$ configuration occurs in the V$^{4+}$ ion in
BaVS$_{3}$, MgV$_{2}$O$_{5}$ \cite{8} or Na$_{2}$V$_{3}$O$_{7}$
\cite{9,10,11}, for instance, or in the Ti$^{3+}$ ion in
LaTiO$_{3}$ or YTiO$_{3}$. In all these compounds 3$d$ cation
sits inside an oxygen/sulphur octahedron, slightly distorted in
the reality. From Fig. 2 one can see that even for the purely
octahedral symmetry 6 states in the spin-orbital space (= 3
orbital states in the orbital state) are split in the reality by
the always present on-site spin-orbit coupling, Fig. 2(2). The
scheme 2(1) is equivalent to Fig. 1b ($O_{h}$) of the commented
paper though instead of the orbital notation we mark states by
their symmetry showing also the related spin.

According to us this electronic structure is seen in experiments
like Electron Spin Resonance (ESR) and is a reason for, for
instance, anomalous temperature dependence of the paramagnetic
susceptibility (violation of the Curie law) and of the heat
capacity at low temperatures. Such the electronic structure of
the V$^{4+}$ ion explains \cite{9,10,11} the dramatic reduction,
by factor 9, of the effective moment in Na$_{2}$V$_{3}$O$_{7}$
observed \cite{12} with decrease of temperature from 100 K to 5 K
(\cite{12}.

In Fig. 3 we present the electronic structure of the 3$d^{6}$
configuration occurring in the Fe$^{2+}$ ion in a
slightly-trigonally distorted bromium octahedron in FeBr$_{2}$
(\cite{13}. The local Fe$^{2+}$ ion surroundings is exactly as it
is shown in Fig. 1a right - in FeBr$_{2}$ the shown $z$ axis is
along the hexagonal $c$ axis. The Co$^{3+}$ ion is isoelectronic
to the Fe$^{2+}$ ion and have exactly the same states. The
structure presented in Fig. 3 shows only the splitting of the two
Hund's rules ground term $^{5}D$ that is 25-fold degenerated.
Fig. 3c shows the electronic structure in case of the purely
octahedral CEF in the presence of the always-present spin-orbit
coupling. Fig. 3e presents the effect of the stretched trigonal
distortion - it has been found to occur in FeBr$_{2}$ \cite{13},
that orders magnetically at T$_{N}$ = 14.2~K.  The
doublet-singlet splitting has been found experimentally to amount
to 0.0028 eV only \cite{13}. The case of Fig. 3d occurs for the
trigonally compressed octahedron, i.e. exactly the situation
shown in Fig. 1b ($D_{3d}$) of the commented paper. Just {\bf our
Fig. 3d should be compared with the situation presented in the
commented paper for the $D_{3d}$ symmetry on Fig. 1b}. Everybody
can see enormous difference between these two electronic
structures being a reason for this Comment. Surely instead of 5
orbital states shown in the commented paper there are 25 states
(in fact, 210 states!!!). The lowest 15 states, originating from
the $^{5}T_{2g}$ cubic subterm, are of the fundamental importance
for thermodynamic properties as they can be thermally populated.
In the Co$^{3+}$/Fe$^{2+}$ ion there are 15 atomic-like states
below 0.1 eV. These states are subsequently populated with the
increasing temperature. The scheme shown in Fig. 3d has been
observed for LaCoO$_{3}$ as the excited subterm (due to the
relatively strong octahedral CEF interactions the ground state
becomes $^{1}A_{1}$ originating from $^{1}I$ atomic term
\cite{6}). The experimentally found splitting between the
singlet-doublet states, that is the effect of the trigonal
distortion, amounts to 7.05~K
(=4.9~cm$^{-1}$~=~0.6~meV~=~0.0006~eV) only, \cite{6} and
references within. It is amazing that so small energy difference
between electronic states exists in a solid!!!

For completeness we have to add that 25 states shown in Fig. 3
are only a small part of the full ionic electronic structure of
the Co$^{3+}$/Fe$^{2+}$ ion, that accounts in total 210 states
grouped in 16 atomic terms \cite{14,15}. The effect of the
octahedral CEF interactions on these 16 terms have been
calculated by Tanabe and Sugano already 50 years ago \cite{15}.
These Tanabe-Sugano diagrams have been somehow forgotten in the
modern solid state theory, likely due to an erroneous conviction
that these states are not relevant to solid materials. Similarly,
many-electron CEF states have been used for interpretation of ESR
spectra on diluted systems \cite{16}, but not to a solid, where a
transition-metal cation was the full part of a crystalline
lattice. However, we would like to mention that a quite similar
picture, though schematic, to calculated by us Fig. 3 has been
presented by, for instance, Birgeneau et al. \cite{17} in Phys.
Rev. B in 1972 - we do not know reasons for forgetting works on
the crystal field of early Van Vleck, Tanabe and Sugano, of
Birgenau and of many others. Moreover, we cannot understand
reasons for almost rejection of the crystal field from the modern
solid state physics. We are somehow grateful to the scientific
discrimination of Phys. Rev. Editors (M. Blume, Wells, ..), to
Polish scientific institutions, to Prof. Prof. H. Szymczak, J.
Sznajd, J. Klamut, A. M. Oles as their negative opinions are the
best proof for the shortage of knowledge on the (many-electron)
crystal field in the XXI century solid-state theory.

Finally, we would like to mention that an enormous separation
energy between $a_{1g}$ and $e_{g}$ states in Fig. 1b, right,
found in LDA band-structure calculations of 1.6 eV \cite{18} we
consider as totally wrong and it will be a subject of another
Comment. In the commented paper a value for the trigonal
distortion effect of 1 eV, mentioned in Ref. 32 of \cite{1}, is
used - this value is, according to us, at least two orders of
magnitude too large. In the approach of the commented paper this
trigonal distortion is of the same size as the octahedral
splitting (the octahedral CEF parameter 10Dq is taken as 0.5 eV
(HS) and 1.5 eV (LS) \cite{1} in Ref. 32), what we consider as
wrong and being in a sharp conflict with the octahedra
construction of the Na$_{x}$CoO$_{2}$ lattice. Our value
B$_{4}$~=~+200~K corresponds to 10Dq~=~2.06~eV \cite{19}. Recent
data indicate for a value 10Dq of 3.0 eV for LaCoO$_{3}$ \cite{6}.

In conclusion, we claim that the electronic structure considered
in a Phys. Rev. Lett. paper of Wu et al. \cite{1} for the Co ion
in the CoO$_{6}$ octahedron in Na$_{x}$CoO$_{2}$, Fig. 1b, is
completely wrong. For physically adequate electronic structure it
is necessary to take into account strong intra-atomic electron
correlations and spin-orbit coupling. The role of the lattice
distortions is very important but their energies are not of order
of 1 eV as LDA band-structure calculations yield, but rather of
0.05 eV as obtained in the many-electron CEF approach. The
many-electron CEF model, with strong electron correlations,
provides in a very natural way the insulating ground state for
3$d$ oxides. Apart of the ground state it allows for calculations
of thermodynamics.

\section {Appendix A} Description of the trigonal distortion within the crystal-field
theory {\cite{20}. After our paper Phys. Rev. B {\bf63} (2001)
172404 \cite{13}.

The 25 levels, originated from the $^{5}$D term, and their
eigenfunctions have been calculated by the direct diagonalization
of the Hamiltonian (1) within the $|LSL_{z}S_{z}\rangle $ base.
It takes a form:

\begin{equation}
H_{d}=H_{cub}+\lambda LS+B_{0}^{2}O\,_{0}^{2}+\mu
_{B}(L+g_{s}S)B  \eqnum{1}
\end{equation}

The separation of the crystal-electric-field (CEF) Hamiltonian
into the cubic and off-cubic part is made for the illustration
reason as the cubic crystal field is usually very predominant. In
the crystallographic structure of FeBr$_{2}$ the Fe ion is
surrounded by 6 Br ions. Despite of the hexagonal elementary cell
Br ions form the almost octahedral surrounding - it justifies the
dominancy of the octahedral crystal field interactions. Moreover,
this octahedral surrounding in the hexagonal unit cell can be
easily distorted along the local cube diagonal - in the hexagonal
unit cell this local cube diagonal lies along the hexagonal {\it
c} axis. The related distortion can be described as the trigonal
distortion of the local octahedron. The cubic CEF Hamiltonian
takes, for the {\it z} axis along the cube diagonal, the form
\begin{equation}
H_{cub}~~=~~-~\frac{2}{3} B_{4}\cdot
(O_{4}^{0}-20\sqrt{2}O_{4}^{3})
\end{equation}

where $O\,_{m}^{n}$ are the Stevens operators. The last term in
Eq. (1) allows studies of the influence of the magnetic field
\cite{13}.

For remembering, the octahedral CEF Hamiltonian with the $z$ axis
along the cube edge takes a form:

\begin{equation}
H_{d}=B_{4}(O_{4}^{0}+5O_{4}^{4})
\end{equation}

$^\spadesuit$ dedicated to Hans Bethe, Kramers and John H. Van
Vleck, pioneers of the crystal-field theory, to the 75$^{th}$
anniversary of the crystal-field theory, and to the Pope John
Paul II, a man of freedom and honesty in life and in Science.


\begin{thebibliography}{9}
\bibitem{20}Prof. H. Szymczak from the Polish Academy of Sciences, the
present long-lasting chairman of the Physical Department of PAS,
has claimed that the procedure used by us is incorrect what is an
erroneous reproach, see an Abragam-Bleaney text-book, for
instance. We mention this problem here because if so great
physicist as Prof. H. Szymczak is does not know the CEF
description of the trigonal distortion it means that the
knowledge about CEF is rather small.
\end{thebibliography}
\end{document}